\documentstyle[11pt]{article}
\newcommand{\beq}{\begin{equation}}
\newcommand{\eeq}{\end{equation}}
\newcommand{\bea}{\begin{eqnarray}} 
\newcommand{\eea}{\end{eqnarray}}
\newcommand{\bq}{\begin{quote}}
\newcommand{\eq}{\end{quote}}

\def\a{\alpha}
\def\b{\beta}

\def\g{\gamma}

\def\s{\sigma}
\def\D{\Delta}
\def\L{\Lambda}

\def\ot{\otimes}
\def\f|{\left |}
\def\r|{\right |}
\def\<{\left <}
\def\>{\right >}

\def\upa{\uparrow}
\def\doa{\downarrow}
\def\ch{\check}

\def\beq{\begin{equation}}
\def\eeq{\end{equation}}
\def\bea{\begin{eqnarray}}
\def\eea{\end{eqnarray}}
\def\ba{\begin{array}}
\def\ea{\end{array}}
\def\no{\nonumber}

\begin{document}
\begin{titlepage}

\begin{center}
{\Large\bf New R-matrices with non-additive spectral parameters and integrable models of strongly correlated fermions}
\vskip.3in
{\large Yao-Zhong Zhang \footnote{Corresponding Author. Email: yzz@maths.uq.edu.au}  and  Jason L. Werry}
\vskip.1in
{\large\em School of Mathematics and Physics, The University of Queensland, \\\ Brisbane, Qld 4072, Australia}

\end{center}
\vskip.6in
\begin{center}
{\bf Abstract}
\end{center}
We present a general formula for constructing R-matrices with non-additive spectral parameters associated with a type-I quantum superalgebra. The spectral parameters originate from two one-parameter families of inequivalent finite-dimensional irreducible representations of the quantum superalgebra upon which the R-matrix acts. Applying to the quantum superalgebra $U_q(gl(2|1))$, we obtain the explicit expression for the $U_q(gl(2|1))$-invariant R-matrix which is of non-difference form in spectral parameters.
Using this R-matrix we derive a new two-parameter integrable model of strongly correlated electrons with pure imaginary pair hopping terms.




\end{titlepage}

\section{Introduction} Type-I quantum superalgebras are known to admit non-trivial one-parameter families of inequivalent finite dimensional irreducible representations (irreps), even for generic $q$ \cite{Gould95}. 
Using such irreps, R-matrices which depend not only on a spectral parameter but in addition on further continuous parameters were constructed in \cite{Bra94,Del95}.
These extra parameters enter the quantum Yang-Baxter equation (QYBE) in non-additive form. The freedom of having such further parameters opens up new and exciting possibilities. For example, by using the rational R-matrix associated with the 4-dimensional irreps of superalgebra $gl(2|1)$, the supersymmetric $U$ model for correlated electrons was introduced in \cite{Bra95}.  The Bethe ansatz solutions of this model were later derived in \cite{Ramos96,Pfa96}. A two-parameter version was proposed in \cite{Bariev95} and  was shown in \cite{Gould96} to come from the trigonometric R-matrix associated with quantum superalgebra $U_q(gl(2|1))$.

There is a different way of treating the extra parameters in the R-matrices obtained in \cite{Bra94,Del95}. Instead of regarding them as free parameters, one can treat them as spectral parameters. This approach was applied in \cite{Links99} to the $gl(2|1)$-invariant rational R-matrix  to derive another extension of the supersymmetric $U$ model. Similar ideas recently also appeared in the construction of R-matrices associated with the central extension of $su(2|2)$ \cite{Beisert07,Beisert08}.
There the values of the central elements, which parametrize the representations of the centrally extended superalgebra, give rise to spectral parameters.  R-matrices obtained from such a construction are also of non-difference form and have found applications in various areas (see e.g. \cite{Martins07,Aru08,Fro12} and references therein).   

In this paper we implement a similar strategy for the one-parameter family of irreps of a general type-I quantum superalgebra. From the results obtained in \cite{Bra94,Del95}, we deduce a formula for quantum superalgebra invariant R-matrices with non-additive spectral parameters. 
Applying our results to the two 4-dimensional irreps of $U_q(gl(2|1))$, we obtain the explicit expression of a new 36-vertex R-matrix without difference property of the spectral parameters. 

The obtained R-matrices can be used to construct new integrable models of correlated fermions. As the R-matrices are of non-difference form, the actual values of the spectral parameters are important in the corresponding model construction. We will restrict our attention to the homogeneous case in which the parameter of each of the representations is the same at each site.  Then the two-site local Hamiltonian of the integrable model corresponding to bivariate R-matrix $R(\a,\b)$ is obtained via
$
H_{12}\sim \sqrt{-1}\; P\left.\frac{\partial}{\partial \a}R(\a,\b)\right|_{\b=\a},\label{construct1}
$
where $P$ is the graded permutation operator.
Using the $U_q(gl(2|1))$-invariant R-matrix derived in this paper, we obtain the local Hamiltonian in terms of the standard electron creation and destruction operators,
\bea
H_{i,i+1}&=&-\sum_{\s}\left(c^{\dag}_{i\s}c_{i+1\s} +h.c.\right)
e^{\frac{1}{2}(\xi-\s\eta)\,n_{i,-\s}+\frac{1}{2}(\xi+\s\eta)\,n_{i+1,-\s}} \no\\
& &+U \,\left(e^{\sqrt{-1}\,\pi/2}\;c_{i\doa}^{\dag}c_{i\upa}^{\dag}c_{i+1\upa}c_{i+1\doa} +h.c.\right),\label{new-model}    
\eea
where $\xi, \eta$ and $U$ are real parameters and $\s=+~ ({\rm or}~\upa), -~({\rm or}~\doa)$. To our knowledge, this is a new integrable model of correlated electrons with pure imaginary pair hopping terms.

\section{R-matrices with non-additive spectral parameters} 
In this section we deduce from the results in \cite{Bra94,Del95} an expression for solutions of the QYBE with non-additive spectral parameters. 

Let $G$ be a simple type-I Lie superalgebra of rank $r$ with generators 
$\{e_i, f_i, h_i, i=1,\cdots, r\}$ and 
$U_q(G)$ the corresponding quantum superalgebra. We will not give the defining relations for $U_q(G)$ here but mention that it is a ${\bf Z}_2$-graded quasi-triangular Hopf algebra with coproduct
\begin{equation}
\Delta(q^{h_i/2})=q^{h_i/2}\otimes q^{h_i/2}\,,~~~\Delta(a)=a\otimes
  q^{-h_i/2}+q^{h_i/2}\otimes a\,,~~~a=e_i,\; f_i.
\end{equation}
Note that $U_q(G)$ also has an opposite coproduct structure defined by 
$\bar{\Delta}=T\cdot \Delta$, with $T$ being the twist map:
$T(a\otimes b)=(-1)^{[a][b]}b\otimes a\,,~\forall a,b\in U_q(G)$, 
where $[a]\in {\bf Z}_2$ denotes the grading of the element $a$. The multiplication rule for the tensor product is defined by
\begin{equation}\label{gradprod}
(a\otimes b)(c\otimes d)=(-1)^{[b][c]}(ac\otimes bd),\qquad \forall a,b,c ,d\in U_q(G).
\end{equation}
 
Let $\pi_\a$ be a one-parameter family of irreducible representations (irreps) provided by the irreducible $U_q(G)$ module $V(\L_\a)$, with the highest weight $\L_\a$ depending on the parameter $\a$. 
Assume that for any $\a$ the irrep $\pi_\a$ is affinizable, i.e. it can be extended to become a loop representation of the corresponding untwisted quantum affine superalgebra $U_q(\hat{G})$. Let $x\in {\bf C}$ be the associated loop parameter (the spectral parameter), and let ${\cal R}(x|\a,\b)\in {\rm End}(V(\L_\a)\otimes V(\L_\b))$ be an operator (R-matrix) associated with two affinizable irreps $\pi_\a, \pi_\b$ from the one-parameter family.
Let
$$\check{\cal R}(x|\a,\b)=P {\cal R}(x|\a,\b),$$ 
where $P$ is the graded permutation operator in $V(\L_\a)\otimes V(\L_\b)$ such that
\begin{equation}\label{gradperm}
P(v_\alpha\otimes v_\beta)=(-1)^{[v_\alpha][v_\beta]}
  v_\beta\otimes v_\alpha\,,~~
  \forall v_\alpha\in V(\L_\a)\,,~v_\beta\in V(\L_\b).
\end{equation}
Then a solution to the linear equations
\begin{eqnarray}
&&\check{\cal R}(x|\a,\b)\Delta^{\a\b}(a)={\Delta}^{\b\a}
  (a)\check{\cal R}(x|\a,\b)\,,~~~\forall a\in U_q(G),\nonumber\\
&&\check{\cal R}(x|\a,\b)\left (x\pi_{\a}(f_\psi)\otimes \pi_{\b}(q^{h_\psi/2})+
  \pi_\a (q^{-h_\psi/2})\otimes \pi_{\b}(f_\psi)\right )\nonumber\\
&&~~~~~~\qquad\qquad  =\left (x\pi_{\b}(f_\psi)\otimes \pi_{\a}(q^{h_\psi/2})
  +\pi_{\b}(q^{-h_\psi/2})\otimes \pi_{\a}(f_\psi)\right )\check{\cal R}(x|\a,\b),\label{r(x)1}
\end{eqnarray}
satisfies the QYBE in the tensor product module $V(\L_\a)\otimes V(\L_\b)\otimes V(\L_\g)$ of three affinizable irreps from the one-parameter family \cite{Jimbo,Bra90,Del95}:
\bea
&&(I\otimes\check{\cal R}(x|\a,\b))(\check{\cal R}(xy|\a,\g)
  \otimes I)(I\otimes\check{\cal R}(y|\b,\g))\nonumber\\
&&\qquad\qquad\qquad\qquad =(\check{\cal R}(y|\b,\g)\otimes I)(I\otimes \check{\cal R}(xy|\a,\g))
  (\check{\cal R}(x|\a,\b)\otimes I).\label{x-dep-non-additive-ybe}
\eea
In the above, $\Delta^{\a\b}(a)=(\pi_{\a}\otimes \pi_{\b})\Delta(a)$ and $f_\psi, h_\psi$ are elements  associated with the highest root $\psi$ of $U_q(G)$
such that the irreps extend to the loop representation of $U_q(\hat{G})$. 
It was also shown in \cite{Jimbo,Bra90} that the solution to (\ref{r(x)1}) is unique  (up to a scalar normalization factor).
In \cite{Del95}, $\a, \b, \g$ were treated as extra free parameters although they enter in the QYBE in non-additive form.

A systematic method was developed in \cite{Del95} for constructing solutions to (\ref{r(x)1}) which satisfy the QYBE (\ref{x-dep-non-additive-ybe}). 
In this work, we focus on the $x=1$ case, i.e. we fix the parameter $x$ to 1, but instead treat the parameters $\a,\b$ as the spectral parameters. Set 
\beq
\check{R}(\a,\b)\equiv \check{\cal R}(1|\a,\b).\label{x=1}
\eeq
It then follows from (\ref{r(x)1}) and (\ref{x-dep-non-additive-ybe}) that the linear relations
\begin{eqnarray}
&&\check{\cal R}(\a,\b)\Delta^{\a\b}(a)={\Delta}^{\b\a}
  (a)\check{\cal R}(\a,\b)\,,~~~\forall a\in U_q(G),\nonumber\\
&&\check{\cal R}(\a,\b)\left (\pi_{\a}(f_\psi)\otimes \pi_{\b}(q^{h_\psi/2})+
  \pi_\a (q^{-h_\psi/2})\otimes \pi_{\b}(f_\psi)\right )\nonumber\\
&&~~~~~~\qquad\qquad  =\left (\pi_{\b}(f_\psi)\otimes \pi_{\a}(q^{h_\psi/2})
  +\pi_{\b}(q^{-h_\psi/2})\otimes \pi_{\a}(f_\psi)\right )\check{\cal R}(\a,\b),
\label{r2}
\end{eqnarray}
admit a {\em unique} solution $\check{\cal R}(\a,\b)$ (under suitable normalization) for any given one-parameter family of irreps of $U_q(G)$ which satisfies the QYBE 
\beq
(I\otimes\check{R}(\a,\b))(\check{R}(\a,\g)
  \otimes I)(I\otimes\check{R}(\b,\g))\nonumber=(\check{R}(\b,\g)\otimes I)(I\otimes \check{R}(\a,\g))
  (\check{R}(\a,\b)\otimes I)\label{non-additive-ybe}
\eeq
with non-additive spectral parameters $\a,\b,\g$, provided such irreps are consistently affiniable.
   
In the case of a multiplicity-free tensor product decomposition
\begin{equation}
V(\L_\a)\otimes V(\L_\b)=\bigoplus_\mu V(\mu),\label{free}
\end{equation}
where $\mu$ denotes a highest weight depending on the parameters $\a$
and $\b$, solution  $\check{R}(\a,\b)$ of (\ref{r2}) can be written in the general form  \cite{Del95}
\beq
\check{R}(\a,\b)=\sum_{\mu}\;\rho_\mu\;  {\bf P}^{\a\b}_\mu.\label{r-check'}
\eeq
Here $\rho_\mu$ are combination coefficients to be determined and ${\bf P}^{\a\b}_\mu:~V(\L_\a)\otimes V(\L_\b)\rightarrow V(\mu)\subset V(\L_\b)\otimes V(\L_\a)$ are the elementary intertwiners satisfying
\beq
{\bf P}^{\a\b}_\mu\Delta^{\a\b}(a)=\Delta^{\b\a}(a)
{\bf P}^{\a\b}_\mu,\qquad ~  \forall a\in U_q(G).\label{intertwiner}
\eeq 
Explicit expressions for ${\bf P}^{\a\b}_\mu$ can be constructed as follows.
Let $\{|e^\mu_i\rangle_{\a\otimes \b}\}$ be an orthonormal
basis for $V(\mu)$ in $V(\L_\a)\otimes V(\L_\b)$.
$V(\mu)$ is also embedded in $V(\L_\b)\otimes V(\L_\a)$ through the
opposite coproduct $\bar{\Delta}$. Let
$\{|e^\mu_i\rangle_{\b\otimes \a}\}$ denote the corresponding orthonormal basis. 
Using these bases, ${\bf P}^{\a\b}_{\mu}$ can be expressed as \cite{Bra94,Del95}
\beq
{\bf P}^{\a\b}_\mu=\sum_i\left |e^\mu_i\right
  \rangle_{\b\otimes\a~
  \a\otimes\b}\!\left\langle e^\mu_i\right |\,.\label{elem-intertwiner}
\eeq
Obviously, when $\a=\b$, ${\bf P}^{\a\a}_\mu$ are the usual projection operators which satisfy $\sum_\mu\, {\bf P}^{\a\a}_\mu=I$. 
It can be easily shown that ${\bf P}^{\a\b}_{\mu}$ satisfies the relation
\beq
{\bf P}^{\b\a}_\mu\,{\bf P}^{\a\b}_{\mu'}=\delta_{\mu\mu'}
  {\cal P}^{\a\b}_\mu, \label{PPP}
\eeq
where ${\cal P}^{\a\b}_\mu:~V(\L_\a)\otimes V(\L_\b)
\rightarrow V(\mu)$ are the projection operators satisfying $\sum_\mu {\cal P}^{\a\b}_\mu=I$.

Thanks to the uniqueness of the solution to (\ref{r2}), it suffices to consider the case that the coefficients $\rho_\mu$ in (\ref{r-check'}) are independent of the parameters $\a, \b$. To determine such $\rho_\mu$, we normalize $\check{R}(\a,\b)$ as 
\beq
\check{R}(\a,\a)=I,\qquad \check{R}(\a,\b)\check{R}(\b,\a)=I,\label{regularity-unitarity}
\eeq
which are called the regularity and unitarity conditions, respectively. 
Then the unitarity condition $\check{R}(\a,\b)\check{R}(\b,\a)=I$ and the relation (\ref{PPP}) mean that $\rho_\mu$ satisfies $(\rho_\mu)^2=1$, so that $\rho_\mu=\pm 1$. By using the regularity condition $\check{R}(\a,\a)=I$  and  the property $\sum_\mu\, {\bf P}^{\a\a}_\mu=I$, we can conclude that $\rho_\mu$ appearing in (\ref{r-check'}) must  equal to 1 identically. It follows that solution $\check{R}(\a,\b)$ of (\ref{r2}) has the particularly simple form,
\beq
\check{R}(\a,\b)=\sum_{\mu}\;  {\bf P}^{\a\b}_\mu\label{r-check}
\eeq
and this solution satisfies the QYBE (\ref{non-additive-ybe}) with non-difference property of the spectral parameters.
 
Let's remark that the simple formula (\ref{r-check}) is essentially $x=1$ specialization of the R-matrix in \cite{Del95} associated with the affine $U_q(\hat{G})$. It is different from the standard R-matrix from the non-affine $U_q(G)$ (which corresponds to the $x=0$ specialization of the R-matrix in \cite{Del95}).  

\section{Explicit expression for $U_q(gl(2|1))$-invariant new R-matrix}

We now apply the above formalism to the one-parameter family of 4-dimensional irreps of the quantum superalgebra $U_q(gl(2|1))$ with hight weight $\L_\a=(0,0|\a)$.  It is known that such irreps of $U_q(gl(2|1))$ are affinizable and thus the formula (\ref{r-check}) can be applied to construct the corresponding R-matrix which satisfies the QYBE (\ref{non-additive-ybe}).

$U_q(gl(2|1))$ has generators $\{E^i_j\}_{i,j=1}^3$ with grading determined by $\left [E^i_j\right ]=([i]+[j])\;(\rm mod\,\,2)$, where $[1]=[2]=0,\,[3]=1$.
Let $\{\f|v\>\}_{v=1}^4$ denote an orthonormal 
basis for a 4-dimensional $U_q(gl(2|1))$ module $V(\L_\a)$. Consistent with the ${\bf Z}_2$-grading of the generators we may assign a grading on the basis states  by
\beq
[\f|1\>]=[\f|4\>]=0,\qquad[\f|2\>]=[\f|3\>]=1.   \label {grading} 
\eeq
Then the generators $\{E^i_j\}_{i,j=1}^3$ act on this module according to  
\begin{eqnarray}
E^1_2&=&\f|2\>\<3\r|,\quad E^2_1=\f|3\>\<2\r|,\quad E^1_1=-\f|3\>\<3\r|
-\f|4\>\<4\r|,\quad 
E^2_2=-\f|2\>\<2\r|-\f|4\>\<4\r|, \nonumber  \\
E^2_3&=&\sqrt{[\a]_q}\f|1\>\<2\r|+\sqrt{[\a+1]_q}\f|3\>\<4\r|,\quad
E^3_2=\sqrt{[\a]_q}\f|2\>\<1\r|+\sqrt{[\a +1]_q}\f|4\>\<3\r|,\nonumber  \\
E_3^3&=&\a\f|1\>\<1\r|+(\a+1)(\f|2\>\<2\r|+\f|3\>\<3\r|)+
(\a+2)\f|4\>\<4\r|,   \label {rep} 
\end{eqnarray}    
where $[x]_q=(q^x-q^{-x})/(q-q^{-1})$.  
Note that the representation depends upon a parameter $\a\in{\bf C}$. 
For $\a>0$ we have 
$\left (E^i_j\right )^{\dag}=E_i^j $ 
and we call the representation unitary of type I. For $\a<-1$ we have
$\left (E_j^i\right )^{\dag}=(-1)^{[i]+[j]}E_i^j $ 
and we say the representation is unitary of type II. We will assume
that $\a $ is restricted to either of the above ranges. 

Associated with $U_q(gl(2|1))$ there is  a co-product structure
(${\bf Z}_2$-graded algebra homomorphism) $\D:U_q(gl(2|1))\rightarrow
U_q(gl(2|1))\ot U_q(gl(2|1))$ given by
\bea
&&\D(E^i_i)=I\ot E_i^i +E_i^i\ot I, \qquad i=1,2,3 \nonumber  \\
&&\D(E_2^1)=E_2^1\ot q^{-\frac 12(E^1_1-E^2_2)}+q^{\frac
12(E_1^1-E_2^2)}\ot E_2^1, \no\\
&&\D(E_1^2)=E_1^2\ot q^{-\frac 12(E^1_1-E^2_2)}+q^{\frac 12(E_1^1-E_2^2)}\ot E_1^2, \nonumber  \\
&&\D(E_3^2)=E_3^2\ot q^{-\frac 12(E_2^2+E_3^3)}+ q^{\frac 12
(E_2^2+E_3^3)} \ot E_3^2,\no\\
&&\D(E_2^3)=E_2^3\ot q^{-\frac 12(E_2^2+E_3^3)}+ q^{\frac 12
(E_2^2+E_3^3)} \ot E_2^3. \label {coproduct} 
\eea   
Under the co-product action the tensor product $V(0,0|\a)\ot V(0,0|\b)$ is also a $U_q(gl(2|1))$ module which reduces completely;
\beq
V(0,0|\a)\ot V(0,0|\b)=V_1\oplus V_2\oplus V_3, \label{decom}  
\eeq
where $V_1\equiv V(0,0|\a+\b)$ and $V_3\equiv V(-1,-1|\a+\b+2)$ are 4-dimensional modules and $V_2\equiv V(0,-1|\a+\b+1)$ is
8-dimensional. 

Let ${\bf P}_k^{\a\b},\,\,k=1,2,3$ denote the intertwining operators from
$V(0,0|\a)\ot V(0,0|\b)$ onto $V_k\in V(0,0|\b)\ot V(0,0|\a)$. Then 
the intertwining operators may be evaluated as follows. 

Let $\f|\Psi^1_k\>_{\a\ot\b},
\,\f|\Psi^3_k\>_{\a\ot\b},\,k=1,2,3,4$ and $\f|\Psi^2_l\>_{\a\ot\b}, \,l=1,2,\cdots,8$ form symmetry adapted bases for the spaces
$V_1, V_3$ and $V_2$ respectively. By means of the representation (\ref{rep}) and the co-product action (\ref{coproduct}) we obtain
\begin{eqnarray}
|\Psi^1_1 \rangle_{\alpha \otimes \beta} &=&  |1\rangle \otimes  |1\rangle,\no \\
|\Psi^1_2 \rangle_{\alpha \otimes \beta} &=& \frac{1}{\sqrt{[\alpha + \beta]_q}} \left[ q^{-\beta/2} \sqrt{[\alpha]_q}\, (|2\rangle \otimes |1\rangle) + q^{\alpha/2} \sqrt{[\beta]_q}\,(|1\rangle \otimes |2\rangle)  \right],\no \\
|\Psi^1_3 \rangle_{\alpha \otimes \beta} &=& \frac{1}{\sqrt{[\alpha + \beta]_q}} \left[ q^{-\beta/2} \sqrt{[\alpha]_q}\, (|3\rangle \otimes |1\rangle) + q^{\alpha/2} \sqrt{[\beta]_q}\,(|1\rangle \otimes |3\rangle)  \right],\no \\
|\Psi^1_4 \rangle_{\alpha \otimes \beta} &=& \frac{1}{\sqrt{[\alpha + \beta]_q [\alpha + \beta + 1]_q}} \{ q^{-\beta} \sqrt{[\alpha]_q [\alpha+1]_q}\,  (|4\rangle \otimes |1\rangle) + q^{\alpha} \sqrt{[\beta]_q [\beta + 1]_q}\, (|1\rangle \otimes |4\rangle)\},\no  \\
& &~~~+ \sqrt{[\alpha]_q[\beta]_q }\,\left(q^{(\alpha - \beta - 1)/2} |2\rangle \otimes |3\rangle - q^{( \alpha - \beta + 1)/2} |3\rangle \otimes |2\rangle \right) \},\no \\
|\Psi^2_1 \rangle_{\alpha \otimes \beta} &=& \frac{1}{\sqrt{[\alpha + \beta]_q}} \left[ q^{\alpha/2} \sqrt{[\beta]_q}\, (|2\rangle \otimes |1\rangle) - q^{-\beta/2}\sqrt{[\alpha]_q}\,(|1\rangle \otimes |2\rangle)  \right],\no \\
|\Psi^2_2 \rangle_{\alpha \otimes \beta} &=&  \frac{1}{\sqrt{[\alpha + \beta]_q}} \left[ q^{\alpha/2} \sqrt{[\beta]_q}\, (|3\rangle \otimes |1\rangle) - q^{-\beta/2}\sqrt{[\alpha]_q}\,(|1\rangle \otimes |3\rangle)  \right],\no \\
|\Psi^2_3 \rangle_{\alpha \otimes \beta} &=& \frac{1}{\sqrt{[\alpha + \beta]_q [\alpha + \beta + 1]_q}} \{ q^{(\alpha - \beta)/2} \sqrt{[\alpha + 1]_q [\beta]_q}\, (|4\rangle \otimes |1\rangle),\no\\
& & - q^{(\alpha -\beta)/2} \sqrt{[\alpha]_q [\beta + 1]_q}\, (|1\rangle \otimes |4\rangle),\no\\
& & - q^{-\beta  - 1/2} [\alpha]_q\,(|2\rangle \otimes |3\rangle) -  q^{\alpha + 1/2} [\beta]_q\, (|3\rangle \otimes |2\rangle) \},\no\\
|\Psi^2_4 \rangle_{\alpha \otimes \beta} &=& |3\rangle \otimes |3\rangle,\no \\
|\Psi^2_5 \rangle_{\alpha \otimes \beta} &=& \frac{1}{\sqrt{[\alpha + \beta + 2]_q }}\left( q^{-(\beta + 1)/2} \sqrt{[\alpha + 1]_q}\, (|4\rangle \otimes |3\rangle) - q^{(\alpha+1)/2} \sqrt{[\beta+1]_q}\,(|3\rangle \otimes |4\rangle) \right),\no \\
|\Psi^2_6 \rangle_{\alpha \otimes \beta} &=& \frac{1}{\sqrt{[\alpha + \beta + 2]_q }}\left( q^{-(\beta + 1)/2} \sqrt{[\alpha + 1]_q}\, (|4\rangle \otimes |2\rangle) - q^{(\alpha+1)/2} \sqrt{[\beta+1]_q}\,(|2\rangle \otimes |4\rangle) \right),\no \\
|\Psi^2_7 \rangle_{\alpha \otimes \beta} &=& \frac{1}{\sqrt{ [\alpha + \beta + 1]_q [\alpha + \beta + 2]_q}}\{~ q^{(\alpha - \beta)/2} \sqrt{[\alpha + 1]_q [\beta]_q }\,(|4\rangle \otimes |1\rangle),\no \\
& &- q^{(\alpha - \beta)/2} \sqrt{[\beta+1]_q [\alpha]_q}\, (|1\rangle \otimes |4\rangle)+ q^{(\alpha+1/2)}[\beta+1]_q\,(|2\rangle \otimes |3\rangle),\no \\
& &\qquad + q^{-(\beta + 1/2)} [\alpha + 1]_q \,(|3\rangle \otimes |2\rangle)   \},\no \\
|\Psi^2_8 \rangle_{\alpha \otimes \beta} &=& |2\rangle \otimes |2\rangle,\no \\
|\Psi^3_1 \rangle_{\alpha \otimes \beta} &=& \frac{1}{\sqrt{[\alpha + \beta + 1]_q [\alpha + \beta + 2]_q}} \{ q^{(\alpha-\beta+1)/2}\sqrt{[\beta + 1]_q [\alpha + 1]_q}\, (|3\rangle \otimes |2\rangle),\no\\
& &- q^{-(\beta - \alpha + 1)/2}\sqrt{[\alpha + 1]_q [\beta + 1]_q}\, (|2\rangle \otimes |3\rangle),\no\\
& & + q^{\alpha+1} \sqrt{[\beta + 1]_q [\beta]_q}\,(|4\rangle \otimes |1\rangle) + q^{-\beta - 1}\sqrt{[\alpha + 1]_q [\alpha]_q}\, (|1\rangle \otimes |4\rangle) \} ,\no\\
|\Psi^3_2 \rangle_{\alpha \otimes \beta} &=& \frac{1}{\sqrt{[ \alpha + \beta + 2 ]_q }}\left( q^{(\alpha+1)/2} \sqrt{[\beta + 1]_q}\,(|4\rangle \otimes |2\rangle)  + q^{-(\beta + 1)/2}\sqrt{[\alpha + 1]_q }\,(|2\rangle \otimes |4\rangle) \right),\no\\
|\Psi^3_3 \rangle_{\alpha \otimes \beta} &=& \frac{1}{\sqrt{[ \alpha + \beta + 2 ]_q }}\left( q^{(\alpha+1)/2} \sqrt{[\beta + 1]_q}\,(|4\rangle \otimes |3\rangle)  + q^{-(\beta + 1)/2}\sqrt{[\alpha + 1]_q }\,(|3\rangle \otimes |4\rangle) \right),\no\\
|\Psi^3_4 \rangle_{\alpha \otimes \beta} &=& |4\rangle \otimes |4\rangle.\label{basis}
\end{eqnarray}
The dual vectors of these basis vectors are defined using the following rules, 
\bea
&&\left (\f|x\>\ot\f|y\>\right)^{\dag}=(-1)^{[|x>][|y>]}\<x\r|\ot\<y\r|,\no\\
&&\<\Psi_k^i\r|_{\a\ot\b}=\f|\Psi^i_k\>^{\dag}_{\a\ot\b},\qquad i=1,2,3.
\eea
It can be checked that (\ref{basis}) forms an orthonormal basis \footnote{A set of non mutually orthogonal basis vectors for $V_i$ in the decomposition (\ref{decom}) was obtained in \cite{DGZ-unpublished}.}. 
It then follows from (\ref{r-check}) and (\ref{elem-intertwiner}) that the $U_q(gl(2|1))$-invariant R-matrix depending on the spectral parameters $\a, \b$ is given by
\bea
\ch R(\a,\b)&=&{\bf P}_1^{\a\b}+{\bf P}_2^{\a\b}+{\bf P}_3^{\a\b},\\
{\bf P}_i^{\a\b}&=&\sum_{k}\f|\Psi^i_k\>_{\b\ot\a~\a\ot\b}\<\Psi_k^i\r|,\qquad 
i=1,2,3.
\eea
By means of the orthonormal basis vectors (\ref{basis}) we obtain the following explicit expression for the $U_q(gl(2|1))$-invariant R-matrix which satisfies the QYBE (\ref{non-additive-ybe}),  
\bea
\check{R}(\alpha,\beta) &=& 
e_{11} \otimes e_{11} 
+e_{22} \otimes e_{22}
+e_{33} \otimes e_{33} 
+e_{44} \otimes e_{44} \no\\
& &+ \left(q^{(\alpha + \beta)/2} + q^{-(\alpha + \beta)/2}\right)\frac{\sqrt{ [\alpha]_q[\beta]_q}}{[\alpha + \beta]_q} (
e_{22} \otimes e_{11} 
+e_{11} \otimes e_{22} 
+e_{33} \otimes e_{11} 
+e_{11} \otimes e_{33} 
) \no\\
& &+ f\frac{\sqrt{[\alpha]_q [\beta]_q  [\alpha + 1]_q [\beta + 1]_q  }}{[\alpha + \beta]_q[\alpha + \beta + 2]_q} (
e_{44} \otimes e_{11} 
+e_{11} \otimes e_{44} 
)\no\\
& &+ q^{-1} \frac{ f [\alpha]_q[\beta]_q + \left( q^{\alpha + \beta + 2} + 1 \right)[\alpha + \beta]_q }{[\alpha + \beta]_q[\alpha + \beta + 2]_q} e_{22} \otimes e_{33} \no\\
& &+ q \frac{ f [\alpha]_q[\beta]_q + \left( q^{-\alpha - \beta - 2} + 1 \right)[\alpha + \beta]_q }{[\alpha + \beta]_q[\alpha + \beta + 2]_q} e_{33} \otimes e_{22} \no\\
& &+ \left(q^{-(\alpha + \beta + 2)/2} + q^{(\alpha + \beta + 2)/2} \right)\frac{\sqrt{ [\alpha + 1]_q[\beta + 1]_q} }{[ \alpha + \beta + 2 ]_q}  \no\\
& & \qquad\qquad\qquad \qquad\qquad\qquad~~\times (
e_{44} \otimes e_{22} 
+e_{22} \otimes e_{44} 
+e_{44} \otimes e_{33} 
+e_{33} \otimes e_{44} 
)\no\\
& &+ \frac{[\beta]_q - [\alpha]_q}{[\alpha + \beta]_q} (
e_{21} \otimes e_{12} 
+e_{12} \otimes e_{21} 
+e_{31} \otimes e_{13} 
+e_{13} \otimes e_{31} 
)\no\\
& &+ f\frac{[\frac{\alpha -\beta}{2}]_q   }{[\alpha + \beta]_q[\alpha + \beta + 2]_q} \left(\left[\frac{\alpha - \beta - 2}{2}\right]_q e_{41} \otimes e_{14} 
+\left[\frac{\alpha - \beta + 2}{2}\right]_q e_{14} \otimes e_{41} 
\right.\no\\
& &\left.\qquad\qquad\qquad\qquad\qquad\qquad + 
\left[\frac{\alpha - \beta}{2}\right]_q(
e_{23} \otimes e_{32} 
+e_{32} \otimes e_{23} ) 
\right)\no\\
& &+ \frac{[\beta + 1]_q - [\alpha + 1]_q}{[ \alpha + \beta + 2 ]_q}
(
e_{42} \otimes e_{24} 
+e_{24} \otimes e_{42} 
+e_{43} \otimes e_{34} 
+e_{34} \otimes e_{43} 
)\no \\
& &+ f \frac{ [\frac{\alpha - \beta}{2}]_q \sqrt{[\alpha]_q [\beta + 1]_q  }}{[\alpha + \beta]_q[\alpha + \beta + 2]_q} \left(q^{-1/2}  (
e_{42} \otimes e_{13} 
-e_{21} \otimes e_{34}  
)
+ q^{1/2}  (
e_{31} \otimes e_{24} 
- e_{43} \otimes e_{12} 
) \right) \no\\
& &+ f \frac{ [\frac{\alpha - \beta}{2}]_q \sqrt{[\alpha + 1]_q [\beta]_q  }}{[\alpha + \beta]_q[\alpha + \beta + 2]_q}\left(q^{-1/2}  (
e_{24} \otimes e_{31}
-e_{12} \otimes e_{43}   
)
+ q^{1/2} (
e_{13} \otimes e_{42}
-e_{34} \otimes e_{21} 
)\right),\no\\ \label{new-r-matrix}
\eea
where we have used the notation $e_{ij}\equiv \f|i\>\<j\r|$ and 
$$f = q^{\alpha+\beta+1} + q^{-\alpha-\beta-1} + q + q^{-1}.$$
It can be checked that the R-matrix (\ref{new-r-matrix}) satisfies the regularity and unitarity properties (\ref{regularity-unitarity}) as well as the QYBE (\ref{non-additive-ybe}), as required. To our knowledge, the bivariate $U_q(gl(2|1))$-invariant R-matrix (\ref{new-r-matrix}) is new. 

In the $q\rightarrow 1$ limit, we obtain its ``rational" version,
\bea
\check{\bar R}(\alpha,\beta) &=& 
e_{11} \otimes e_{11} 
+e_{22} \otimes e_{22}
+e_{33} \otimes e_{33} 
+e_{44} \otimes e_{44} \no\\
& &+ 2\frac{\sqrt{ \alpha\beta}}{\alpha + \beta} (
e_{22} \otimes e_{11} 
+e_{11} \otimes e_{22} 
+e_{33} \otimes e_{11} 
+e_{11} \otimes e_{33} 
) \no\\
& &+ 4\frac{\sqrt{\alpha\beta(\alpha + 1)(\beta + 1)}}{(\alpha + \beta)(\alpha + \beta + 2)} (
e_{44} \otimes e_{11} 
+e_{11} \otimes e_{44} 
)\no\\
& &+ 2 \frac{\a+ 2\alpha\beta + \beta  }{(\alpha + \beta)(\alpha + \beta + 2)} (e_{22} \otimes e_{33}+  e_{33} \otimes e_{22}) \no\\
& &+ 2\frac{\sqrt{ (\alpha + 1)(\beta + 1)} }{\alpha + \beta + 2 }  (
e_{44} \otimes e_{22} 
+e_{22} \otimes e_{44} 
+e_{44} \otimes e_{33} 
+e_{33} \otimes e_{44} 
)\no\\
& &+ \frac{\beta- \alpha}{\alpha + \beta} (
e_{21} \otimes e_{12} 
+e_{12} \otimes e_{21} 
+e_{31} \otimes e_{13} 
+e_{13} \otimes e_{31} 
)\no\\
& &+ \frac{\alpha -\beta}{(\alpha + \beta)(\alpha + \beta + 2)} \left\{(\alpha - \beta - 2)\, e_{41} \otimes e_{14} 
+(\alpha - \beta + 2)\, e_{14} \otimes e_{41} 
\right.\no\\
& &\left. \qquad\qquad\qquad\qquad\qquad + (\alpha -\beta)(
e_{23} \otimes e_{32} 
+e_{32} \otimes e_{23} 
) \right\}\no\\
& &+ \frac{\beta  - \alpha }{ \alpha + \beta + 2 }
(
e_{42} \otimes e_{24} 
+e_{24} \otimes e_{42} 
+e_{43} \otimes e_{34} 
+e_{34} \otimes e_{43} 
)\no \\
& &+ 2 \frac{ (\alpha - \beta)\, \sqrt{\alpha(\beta + 1) }}{(\alpha + \beta)(\alpha + \beta + 2)} \left(
e_{42} \otimes e_{13} 
-e_{21} \otimes e_{34}  
+ e_{31} \otimes e_{24} 
- e_{43} \otimes e_{12} 
\right) \no\\
& &+ 2 \frac{ (\alpha - \beta)\, \sqrt{(\alpha + 1)\beta }}{(\alpha + \beta)(\alpha + \beta + 2)}\left(
e_{24} \otimes e_{31}
-e_{12} \otimes e_{43}   
+ e_{13} \otimes e_{42}
-e_{34} \otimes e_{21} 
\right). \label{rational-r-matrix}
\eea
This R-matrix is $gl(2|1)$ invariant and also seems new.

\section{New integrable model of strongly correlated electrons}
The new R-matrix (\ref{new-r-matrix}) can be used to define an integrable model of correlated electrons.
On the $N$-fold tensor product space we denote 
\beq
\ch R(\a,\b)_{i,i+1}=I^{\ot (i-1)}\ot\ch R(\a,\b)\ot I^{(N-i-1)},   
\eeq
and define a local Hamiltonian by   
\beq
H_{i,i+1}=-\sqrt{-1}\,\frac{(q^{\a+1}-q^{-\a-1})}{\ln\,q}\left.\frac {\partial}{\partial\a} \ch R(\a,\b)_{i,i+1}\right |_{\b=\a}.  
\eeq
{}From the explicit expression of the R-matrix (\ref{new-r-matrix}) and via a direct computation, we find the two-site local Hamiltonian
\bea
H_{12}&=&\sqrt{-1}\;\frac{[\a+1]_q}{[\a]_q}\,(e_{12}\ot e_{21}+e_{21}\ot e_{12}+e_{13}\ot e_{31}+e_{31}\ot e_{13})\no\\
& &\sqrt{-1}\;\frac{1}{[\a]_q}\,(e_{41}\ot e_{14}-e_{14}\ot e_{41})\no\\
& &+\sqrt{-1}\;(e_{24}\ot e_{42}+e_{42}\ot e_{24}
     +e_{34}\ot e_{43}+e_{43}\ot e_{34})\no\\
& &+\sqrt{-1}\;{\rm sign}(\a)\;\sqrt{\frac{[\a+1]_q}{[\a]_q}}\,\left\{q^{-1/2}\,(e_{12}\ot e_{43} +e_{21}\ot e_{34}-e_{42}\ot e_{13}-e_{24}\ot e_{31})\right.\no\\
& &\qquad\qquad\qquad\qquad \left.+q^{1/2}\,(e_{34}\ot e_{21}+e_{43}\ot e_{12}-e_{13}\ot e_{42}
   -e_{31}\ot e_{24})\right\}.
\eea
In view of the grading (\ref{grading}) we now make the assignment
\beq
\f|4\>\equiv \f|0\>,\qquad \f|3\>\equiv \f|\upa\>=c^{\dag}_{\upa}\f|0\>,\qquad
\f|2\>\equiv \f|\doa\>=c^{\dag}_{\doa}\f|0\>, \qquad \f|1\>\equiv
\f|\upa\doa\>=c_{\doa}^{\dag}c_{\upa}^{\dag}\f|0\>,  
\eeq   
which allows us to express $e_{ij}\equiv \f|i\>\<j\r|$ in terms of the
canonical fermion operators. We can show that we have
\bea
H_{i,i+1}&=&-\sum_{\s}\left(e^{-\sqrt{-1}\,\pi/2}\, c^{\dag}_{i\s}c_{i+1\s} +h.c.\right)\left\{ 1-n_{
i,-\s}\left(1+{\rm sign}(\a)\;  q^{-\s/2}\;\sqrt{\frac {[\a+1]_q}{[\a]_q}}
\right)\right.     \no\\   
& &\qquad\qquad -n_{i+1,-\s}\left(1+{\rm sign}(\a)\; q^{\s/2}\;\sqrt{\frac{[\a+1]_q}{[\a]_q}}\right) \no \\ 
& & \qquad    \left. +n_{i,-\s}n_{i+1,-\s}\left(1+\frac{[\a+1]_q}{[\a]_q} 
 +{\rm sign}(\a)\; (q^{1/2} +q^{-1/2})\;\sqrt{\frac{
[\a+1]_q}{[\a]_q}}\right) \right \}  \no \\    
& &  +\frac{1}{[\a]_q}\left(e^{-\sqrt{-1}\,\pi/2}\,c_{i\doa}^{\dag}c_{i\upa}^{\dag}c_{i+1\upa}c_{i+1\doa} +h.c.\right). \label{local-H}   
\eea
Here and throughout, we have used the standard notation for electron spins: $\s=+$ (or $\upa$), $-$ (or $\doa$). Note that the local Hamiltonian is hermitian only for $\a>0$ or $\a<-1$; i.e. when the underlying representation is unitary.

Under the unitary transformation
\beq
c_{i\s}\rightarrow c_{i\s}(1-2n_{i,-\s})
\eeq
we obtain the same local Hamiltonian with ${\rm sign}(\a)$ replaced by $-{\rm sign}(\a)$. This unitary transformation allow us write (\ref{local-H}) in a compact form. Indeed, after performing the further unitary transformation $c_{i\s}\rightarrow e^{\sqrt{-1}\,i\pi/2}c_{i\s}$, we derive from (\ref{local-H}) the local Hamiltonian (\ref{new-model}) of our new electron model with the parameters $\xi, \eta$ and $U$ defined by 
\beq
e^\eta=q,\qquad e^\xi=\frac{[\a+1]_q}{[\a]_q},\qquad U=\frac{1}{[\a]_q}.\label{xi-eta-U}
\eeq
On a periodic lattice, the corresponding global Hamiltonian of our electron model (which is integrable with periodic boundary conditions) reads 
\bea
H=\sum_i\, H_{i,i+1}&=&-\sum_{i,\s}\left(c^{\dag}_{i\s}c_{i+1\s} +h.c.\right)
e^{\frac{1}{2}(\xi-\s\eta)\,n_{i,-\s}+\frac{1}{2}(\xi+\s\eta)\,n_{i+1,-\s}}\no\\ 
& &  +U\sum_{i} \left(e^{\sqrt{-1}\pi/2}c_{i\doa}^{\dag}c_{i\upa}^{\dag}c_{i+1\upa}c_{i+1\doa} +h.c.\right) .\label{new-model-global}    
\eea 

Observing that $\xi, \eta$ and $U$ are related to two independent parameters $\a$ and $q$ through (\ref{xi-eta-U}) above, we can derive two distinct and non-trivial special cases corresponding to the following limits of $q$ and $\a$ values, respectively. In the limit $q\rightarrow 1$ so that $\eta=0$, our model (\ref{new-model-global}) becomes
\beq
H_1=-\sum_{i,\s}\left(c^{\dag}_{i\s}c_{i+1\s} +h.c.\right)
e^{\frac{1}{2}\bar{\xi}\,(n_{i,-\s}+n_{i+1,-\s})} +\bar{U}\sum_{i} \left(e^{\sqrt{-1}\pi/2}c_{i\doa}^{\dag}c_{i\upa}^{\dag}c_{i+1\upa}c_{i+1\doa} +h.c.\right),\label{limit1}    
\eeq
where 
\beq
e^{\bar{\xi}}=\frac{\a+1}{\a},\qquad ~~ \bar{U}=\frac{1}{\a}.
\eeq
Obviously this special case gives an non-trivial integrable electron model with pure imaginary pair hopping terms. It can be checked that this Hamiltonian is also obtainable directly from the R-matrix (\ref{rational-r-matrix}) which is the $q\rightarrow 1$ limit of (\ref{new-r-matrix}).  

On the other hand, in the limit $\a\rightarrow \infty$ so that $\xi=\eta$,  our model (\ref{new-model-global}) reduces to the Bariev model \cite{Bariev91} whose Hamiltonian on the periodic lattice can be written in our notation as
\beq
H_2=-\sum_{i,\s}\left(c^{\dag}_{i\s}c_{i+1\s} +h.c.\right)
e^{\frac{1}{2}(1-\s)\,\eta\,n_{i,-\s}+\frac{1}{2}(1+\s)\,\eta\,n_{i+1,-\s}}.\label{bariev}    
\eeq
In this sense, our electron model (\ref{new-model-global}) may be considered as an integrable one-parameter extension of the Bariev model (\ref{bariev}) with extra pure imaginary pair hopping terms. 
We mention in passing that the integrability of the Bariev model was established independently in \cite{Zhou96,Shiro97} by obtaining an appropriate solution of the standard (i.e. non-graded) QYBE (through mapping the Bariev model into the coupled $XY$ chain). Our R-matrix (\ref{new-r-matrix}) for our model (\ref{new-model-global}) is inherently different from the bivariate R-matrices computed in \cite{Zhou96,Shiro97} for the Bariev model.

\section{Conclusions and discussions} We have demonstrated in this paper that new solutions to the QYBE without difference property may be naturally constructed through R-matrices associated with one-parameter family of irreps of type I quantum superalgebras. As an example we have considered one of the simplest cases and derived the explicit expression of the new 36-vertex bivariate R-matrix associated with the 4-dimensional irreps of the quantum superalgebra $U_q(gl(2|1))$.   
Using this $R$-matrix, we have obtained the Hamiltonian of the new two-parameter integrable model of strongly correlated electrons with pure imaginary pair hopping terms. Our model contains two non-trivial and distinct one-parameter electron models as its special cases.

Open problems for future work include solving our new model by means of the (algebraic) Bethe ansatz method, and
constructing and solving the corresponding open chain models with boundary conditions defined by boundary K-matrices using an appropriate modification of Sklyanin's method \cite{Skl88}. Of particular interest is to consider twisted or anti-periodic boundary condition, i.e. a ring of electrons with M{\"o}bius like topological boundary condition, by means of the Cao-Yang-Shi-Wang method for a topological spin ring \cite{Cao13}. 
It is also important to derive explicit bivariate R-matrices associated with higher rank type-I quantum superalgebras and construct the corresponding integrable models of correlated fermions.
Finally it seems that in  \cite{Beisert07,Beisert08}, the bivariate R-matrices for the centrally extended $su(2|2)$ and $U_q(su(2|2))$ superalgebras were derived somewhat by brute force. It would be interesting to investigate whether or not the superalgebra representation theory approach presented in this paper can also be applied to derive the bivariate R-matrices associated with central extensions of certain superalgebras. We hope to examine some of the above problems and present the obtained results elsewhere.

\section*{Acknowledgements} 

YZZ would like to thank Yupeng Wang for enlightening discussions and kind hospitality at Institute of Physics, Beijing in April 2019, where part of this work started. YZZ also thanks Jon Links for useful comments and suggestions which helped improve the paper. This work was supported by Australian Research Council Discovery Project DP190101529 and National Natural Science Foundation of China (Grant No. 11775177).


\end{document}